\documentclass[pra,aps,twocolumn]{revtex4}
\usepackage{amssymb}
\usepackage{amsmath}
\usepackage{ams}
\usepackage{epsfig}
\usepackage{t1enc}
\usepackage[cp1250]{inputenc}

\begin{document}

\title{Why devil plays dice?}

\author{Andrzej Dragan\footnote{e-mail: dragan@fuw.edu.pl}}

\affiliation{Institute of Theoretical Physics, University of
Warsaw, Ho\.{z}a 69, 00-681 Warsaw, Poland}

\begin{abstract}
Principle of Relativity involving all, not only subluminal,
inertial frames leads to the disturbance of causal laws in a way
known from the fundamental postulates of Quantum Theory. We show
how quantum indeterminacy based on complex probability amplitudes
with superposition principle emerges from Special Relativity.
\end{abstract}

\maketitle

\section{Introduction}

Experiments devised to test Bell's theorem \cite{Bell1964}
indicate that the fundamental laws of physics can't be formulated
using local and deterministic mode of description. According to
Einstein who disbelieved the fundamental meaning of Quantum
Theory, and his famous metaphor the dice are indeed being played
by someone. By who?

The purpose of this paper is to show that, ironically, the reason
for such a mysterious behavior of Nature originates from a more
fundamental theory - Special Relativity. It is well known that
considering superluminal particles or inertial observers leads to
violations of a causal mode of description. In this paper we show
however, that such considerations do not lead neither to the
possibility of sending superluminal information nor to any acausal
paradoxes but only to the known quantum features, such as
indeterminacy of the result of a single measurement and the
description of motion involving complex amplitudes undergoing
linear superposition.

In Sec.~\ref{sec-noreason} we show that no superluminal
communication is possible even if the tachyons interacting with
matter existed, in Sec.~\ref{sec-relativity} we derive the
transformations for all inertial observers and introduce extended
version of the Principle of Relativity. Sec.~\ref{sec-preferred}
and \ref{sec-superposition} present how quantum description of
motion with complex amplitudes undergoing linear superposition
arises when we account for superluminal observers. In
Sec.~\ref{sec-CPT} we discuss the possibility of existence of
tachyons and Sec.~\ref{Conclusions} concludes the paper. Detailed
mathematical considerations are shifted to Appendices
\ref{proof-fourmomentum} and \ref{proof-wavefunction}.

\section{Reason for acausality\label{sec-noreason}}

\begin{figure}
\begin{center}
\epsfig{file=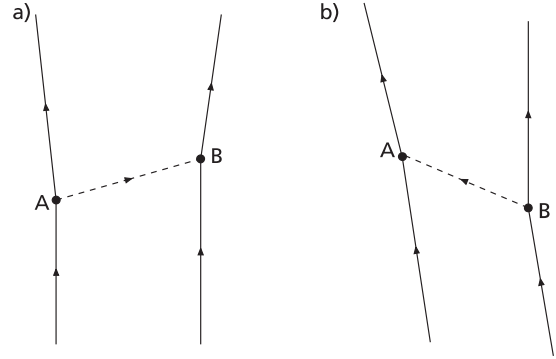} \caption{ \label{Procesy2} \sf
\footnotesize Spacetime diagrams of a process of sending a tachyon
as seen by two inertial observers: a) particle emitted from A and
absorbed in B, b) reversed process in a different inertial frame.}
\end{center}
\end{figure}

Suppose that some local and controllable process is responsible
for the emission of a tachyon with the velocity $w>c$ by a massive
particle at rest - we will denote this event {\sf A} - see
Fig.~\ref{Procesy2}a). After a while the tachyon reaches a
detector located at a distant point - event {\sf B} in
Fig.~\ref{Procesy2}a). Other inertial observer moving with a
relative subluminal velocity $V>c^2/w$ finds out that the time
ordering of the events is opposite - Fig.~\ref{Procesy2}b). He
observes a tachyon emitted by the detector {\sf B} and reaching
the emitter {\sf A} after a while. Let us answer the following
question - what process taking place in the detector {\sf B} in
the second inertial frame could be responsible for the act of
emission of the tachyon? Obviously, no such reason may exist in
the past world-line of the detector {\sf B}, as we assumed that
the process behind the tachyon's emission takes place locally in
{\sf A}. This indicates that in the second inertial frame the act
of emission of the tachyon from the detector {\sf B} is absolutely
{\em spontaneous and deprived of any cause}. Since no frame is
preferred  we deduce that the emission {\sf A} in the first
inertial frame also had to be spontaneous. Our conclusion is that
{\em there is no local, deterministic theory that could describe
emission of a tachyon}. Since it must be a spontaneous process,
{\em no tachyon can be used in superluminal communication},
because the information sent over by a local observer would be
completely out of control. No causal paradoxes arise.

To characterize the process of decay of a "classical" particle
into a given pair of particles one has to specify six components
of the momenta of the products of the decay. There are only four
equations expressing the conservation of energy and momentum, so
the momenta of the products of the decay can't be set uniquely
(with one exception that will be discussed later). It follows from
the analogous reasoning as above that there can be no local
deterministic theory that could determine the momentum of the
emitted tachyon. Its momentum must be therefore attributed
spontaneously.

\section{All inertial observers\label{sec-relativity}}

Consider two inertial frames (unprimed and primed) in a relative
motion with the velocity $\boldsymbol{V}$. Our goal is to
determine the most general form of a transformation of coordinates
between the two frames. The only assumption that we impose is the
Galilean Principle of Relativity \cite{Frank1911}. It follows that
the possible transformation must be linear so that equations do
not distinguish any instant of time or point in space and the
coefficients must be functions of the relative velocity only. From
the Principle of Relativity we also obtain the inversed
transformation. Assuming the relative motion along the common $x$
and $x'$ axis we obtain:
\begin{eqnarray}
\label{lineartransform}
x' &=& A(V) \,x +B(V)\,t,\nonumber \\
x  &=& A(-V)\,x'+B(-V)\,t'.
\end{eqnarray}
From the definition of a relative motion the point $x'=0$ is
described in the unprimed frame by the equation $x=Vt$. Therefore
from the first equation \eqref{lineartransform} we get
$\frac{B(V)}{A(V)} = -V$. Using this identity we can narrow down
the set of possible transformations \eqref{lineartransform} to the
form:
\begin{eqnarray}
\label{lineartransform2}
x' & = & A(V)(x-Vt), \nonumber \\
t'  & = & A(V)\left(t-\frac{A(V)A(-V)-1}{V^2A(V)A(-V)}Vx\right).
\end{eqnarray}
Consider three inertial frames in a relative motion along the $x
\parallel x' \parallel x''$ axis. Let the primed frame move with the
velocity $V_1$ relative to the unprimed frame, and let the bised
frame move with the velocity $V_2$ relative to the primed one. We
determine the transformation between the bised and unprimed system
of coordinates:
\begin{eqnarray}
x'' &=&  A(V_1)A(V_2)x \left(1 + V_1 V_2
\frac{A(V_1)A(-V_1)-1}{V_1^2A(V_1)A(-V_1)} \right)\nonumber \\
& &- A(V_1)A(V_2)(V_1+V_2)t.
\end{eqnarray}
Let us assume that if an object {\sf A} moves with a velocity
$\boldsymbol{V}$ relative to an object {\sf B} then {\sf B} moves
relative to {\sf A} with the velocity $-\boldsymbol{V}$. Therefore
the transformation above should remain unchanged after the
interchange $V_1\leftrightarrow V_2$. Hence we obtain the
condition:
\begin{equation}
\frac{A(V_1)A(-V_1)-1}{V_1^2A(V_1)A(-V_1)}=
\frac{A(V_2)A(-V_2)-1}{V_2^2A(V_2)A(-V_2)}.
\end{equation}
The equality of an unknown function for two arbitrary arguments
$V_1$ and $V_2$ means that the function must be constant:
\begin{equation}
\label{newconstant} \frac{A(V)A(-V)-1}{V^2A(V)A(-V)}=K.
\end{equation}
Consider a frame with a clock with an inversed mechanism, so that
the time flow and all the velocities have the opposite signs. If
the time reversal does not change the spatial coordinates then
from the equation \eqref{lineartransform2} we obtain the condition
$A(-V)=A(V)$ allowing us to determine
$A(V)=\pm\frac{1}{\sqrt{1-KV^2}}$. After the choice of a sign that
guaranties a smooth transition $x'\to x$ when $V\to 0$ we obtain
the final form of the transformation from the equations
\eqref{lineartransform2}:
\begin{eqnarray}
\label{lineartransform3}
x' &=& \frac{x-Vt}{\sqrt{1-KV^2}}, \nonumber \\
t' &=& \frac{t-K V x}{\sqrt{1-KV^2}}.
\end{eqnarray}
The fundamental constant $K$ determining a relation between
spatial dimension $x$ and the temporal dimension $t$ can be equal
to zero, be positive or negative. The first two options correspond
to Galilean and Lorentz transformations, respectively. The
scenario of a negative $K$ describes the world with a
four-dimensional Euclidean space with the fourth dimension $t$
stretched by a factor of $\sqrt{-K}$, and the derived
transformation is just a rotation in the plane $xt$ by the angle
$\tan\alpha = \sqrt{-K}V$. There are four spacetime dimensions
known, coefficients $K$ describing the relations between pairs of
spatial dimensions are all equal to $-1$ and the coefficients
relating time and space are all measured to be equal to $1/c^2$.

To determine the transformation for the perpendicular spatial
direction we note that it must be time-independent. The only
isotropic transformation is therefore of the form $y'=C(V)y$,
$z'=C(V)z$. Let us consider a process of inserting a key into a
keyhole with the velocity $V$. If $|C(V)|<1$ then in the rest
frame of the keyhole the key is perpendicularily contracted and it
can fit in even more easily. However in the key's rest frame the
keyhole is contracted and key can't fit in at all. The same
inconsistency is obtained for $|C(V)|>1$, so we conclude that the
only allowable transformation yields $C(V)=\pm 1$. Since we demand
that for $V\to 0$ the transformation becomes identity, we obtain
$y' = y$ and $z' = z$.

The transformation law \eqref{lineartransform3} is determined
onlyf
 for the subluminal velocities $V<c$. One can however derive
the formulas for the case of superluminal velocities as well. We
will consider the case of an antisymmetric function $A(-V)=-A(V)$.
This assumption leads to the conclusion that the time reversal
$t\to-t$ and consequently $V\to-V$ yield the transformation
$x'\to-x'$ and $t'\to t'$. This follows directly from the
equations \eqref{lineartransform2}. The reason for such a
surprising symmetry law will become clear when we derive the final
form of the equations. From the formula \eqref{newconstant} with
$K=\frac{1}{c^2}$ we obtain $A(W) =
\pm\frac{W/|W|}{\sqrt{W^2/c^2-1}}$ determined for $W>c$ (from now
on we will use $W$'s to denote superluminal velocities and Greek
symbols to denote quantities in superluminal frames). The extra
$W/|W|$ factor is the only antisymmetric function of $W$ of
modulus equal to one. The sign of the function $A(W)$ is not
uniquely determined therefore we obtain \cite{Machildon1983}:
\begin{eqnarray}
\label{lineartransform4}
\chi' &=& \pm\frac{W}{|W|}\frac{x-Wt}{{\sqrt{W^2/c^2-1}}},\nonumber \\
\tau' &=& \pm\frac{W}{|W|}\frac{t-Wx/c^2}{{\sqrt{W^2/c^2-1}}},
\end{eqnarray}
where $\chi'$ is spatial and $\tau'$ temporal dimension related to
the superluminal observer moving with the given velocity $W$. The
last statement is supported by the fact that a temporal axis of a
frame co-moving with a given object must coincide with the
world-line of the object.

As an example example we consider the observer moving with an
infinite velocity along the $x$ axis. It follows that he perceives
the spatial dimension $x$ as the temporal dimension $\tau$ and the
temporal dimension $t$ as the spatial dimension $\chi$ (we choose
the negative sign):
\begin{eqnarray}
\label{lineartransform5}
\chi &=& ct,\nonumber \\
c\tau &=& x.
\end{eqnarray}
This relation justifies the unusual symmetry of the superluminal
transformation discussed previously - the time reversal operation
$t\to-t$ must be related to $\chi\to-\chi$, not $\tau\to-\tau$, as
in the subluminal case. For the two-dimensional spacetime all the
inertial frames including the superluminal ones could be
postulated to be completely undistinguishable by any laws of
physics. In the four-dimensional spacetime, however, the issue is
much more delicate, because of the transformation properties of
the remaining coordinates $y$ and $z$ \cite{Machildon1983}. To
deduce their transformation law we can repeat the same reasoning,
as for the subluminal case. In this case there is no zero-velocity
limit, so the transversal coordinates are defined up to a sign.
Let us denote the four-position of the superluminal observer with
$(\chi', c\tau'_x, c\tau'_y, c\tau'_z)$ and assume the remaining
coordinates to be $c\tau'_y = \pm y$ and $c\tau'_z = \pm z$. Using
these and the equations \eqref{lineartransform4} we derive the
transformation law for the spacetime interval:
\begin{equation}
c^2 \Delta t^2 - \Delta\boldsymbol{r}^2 =
\Delta\chi'^2-c^2\Delta\boldsymbol{\tau}'^2,
\end{equation}
where $\boldsymbol{r}=(x,y,z)$ and
$\boldsymbol{\tau}'=(\tau'_x,\tau'_y,\tau'_z)$. To guarantee the
preservation of the interval we {\em define} the interval in the
superluminal frame as the right-hand side of the above equation.
As we have already pointed out a temporal axis of a frame
co-moving with a given object must coincide with the world-line of
the object, hence $\tau'_x$ must be temporal and $\chi'$ - spatial
coordinate. The nature of the remaining coordinates $\tau'_y$ and
$\tau'_z$ is recognized as temporal dimensions due to their sign
in the metric (the same as $\tau'_x$). The transformations
\eqref{lineartransform4} together with the perpendicular
coordinate transformation is an element of the Lorentz Group,
corresponding to the subluminal velocity $V=c^2/w$ therefore they
preserve the light-cone structure of the spacetime. The
inside-cone four-vectors remain inside the light-cone after an
arbitrary transformation and the outside-cone four-vectors remain
outside. The only new characteristics of the superluminal observer
is that all his time-like four-vectors are (by the definition)
outside-cone vectors and the spatial four-vectors live inside the
cone. The fact that there are three temporal dimensions
$\boldsymbol{\tau}$ and a single spatial dimension $\chi$ will be
discussed later, at this point we only guess that the spacetime
seen from a tachyonic inertial frame has completely different
physical properties from the properties known from subluminal
frames of reference. This seems to essentially limit the
possibility of formulation the Principle of Relativity for all
inertial frames \cite{Machildon1983}, although all subluminal
frames are relativistically equivalent to each other and so are
all the superluminal frames. However we can sustain a weaker
postulate, necessary in any scheme involving the concept of
spacetime. The postulated version of the Principle of Relativity
for all the frames will be stated in the following way: {\em if a
physical process or event takes place in one inertial frame, it
will also take place in any other inertial frame}. The considered
process or event may possibly have quite different properties
according to distinguishable character of the metric in subluminal
and superluminal frames but, the fact that it took place can't
depend on the frame of reference.

The transformation between two superluminal frames can be already
deduced from the reversed transforms between a stationary frame
and two arbitrary superluminal frames. It turns out that such
transformation does not depend on the sign of the transformation
\eqref{lineartransform3}, which shows that the choice of the sign
is, to some degree, only a matter of convention.

Lorentz transformation between sub- and for superluminal frames
has several testable properties, for example a superluminal object
moving with the velocity $w$ along the $x$ axis is observed as
longitudinally distorted in such a way that its length $\Delta x$
equals:
\begin{equation}
\Delta x = \pm\frac{w}{|w|}\Delta\chi\sqrt{w^2/c^2-1},
\end{equation}
where $\Delta\chi$ is the object's stationary length. There is
also a new form of the time flow disturbance of a superluminal
clock:
\begin{eqnarray}
\Delta t  &=& \mp \frac{w}{|w|}\frac{\Delta \tau_x}{\sqrt{w^2/c^2-1}},\nonumber \\
\Delta \tau_y &=& \Delta \tau_z = 0
\end{eqnarray}
so that for $w=\sqrt{2}c$ the length and the time flow are the
same in the stationary and the tachyon's rest frame.

Finally, in Appendix \ref{proof-fourmomentum} we derive and
discuss the simplest candidates for the energy-momentum
four-vector of a tachyon of a mass parameter $\mu$, helicity
$\mbox{\em \'{s}}=\pm 1$ (inevitable in the description) and
velocity $\boldsymbol{w}$:
\begin{equation}
\label{energymomentum}
\begin{split}
E &= \frac{\mbox{\em \'{s}}\,\mu c^2}{\sqrt{w^2/c^2-1}},\\
\boldsymbol{p} &= \frac{\mbox{\em \'{s}}\,\mu \boldsymbol{w}}
{\sqrt{w^2/c^2-1}},
\end{split}
\end{equation}
where the transformation law for $\mbox{\em \'{s}}$ takes the form
$\mbox{\em \'{s}}\,'=\mbox{\em \'{s}} \, \mbox{sgn}
\left(c^2-\boldsymbol{w} \cdot \boldsymbol{V} \right)$. Consider a
decay presented in Fig.~\ref{Procesy3}a), when the decaying
particle reverses its velocity while emitting an infinitely fast
moving tachyon. The tachyon's momentum equals $\mu c$, and the
direction of the emission coincides with the direction of the
velocity of the decaying particle. For given masses $m$ and $\mu$
and the velocity $\boldsymbol{v}$ no other process of decay is
possible - this is the above mentioned exception when the
conservation laws uniquely define the momenta of the products of
the decay. While the momentum is well determined, the position of
the tachyon is completely unknown as it travels with the infinite
velocity. This seems consistent with the conclusions of the
Heisenberg's Principle of Uncertainty.

\section{Preferred scales \label{sec-preferred}}

\begin{figure}
\begin{center}
\epsfig{file=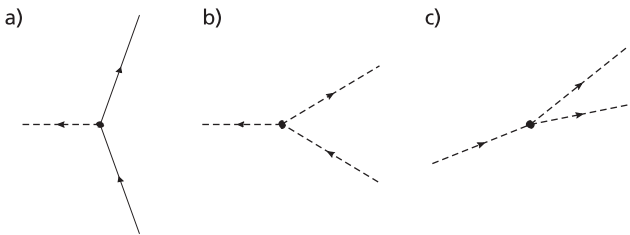} \caption{ \label{Procesy3} \sf
\footnotesize Elastic emission of a tachyon: a) by a massive
particle, b) by another tachyon, c) process b) seen by a different
inertial observer.}
\end{center}
\end{figure}

We have shown that the process of emission shown in
Fig.~\ref{Procesy3}a) can't be described by a local deterministic
theory. The same follows for the process shown in
Fig.~\ref{Procesy3}b) - one can see this by taking the point of
view of another observer - Fig.~\ref{Procesy3}c). It is clear that
it is not possible to attribute to any of the tachyons a hidden
parameter that would govern the process and determine the moment
of its occurrence.

However, according to the transformation \eqref{lineartransform5}
the diagram \ref{Procesy3}b) shows how infinitely fast moving
observer perceives the process of the decay of a massive particle
into a pair of massive particles. From the Principle of Relativity
we conclude that the concept of hidden variables steering the
process of the decay of massive particles cannot be introduced
also in subluminal inertial frames. If there is no local
deterministic parameter in superluminal frames, there cannot be
local deterministic parameters in subluminal frames. Therefore
{\em all the possible processes of decay must be spontaneous.}

The reasoning above agrees with our knowledge of the processes
taking place in the realm of elementary particles but seems to
contradict our experience with macroscopical, massive objects. For
example an ordinary bomb explodes into many pieces at a well
defined instant of time. Obviously the moment of explosion can be
foreseen in advance. We propose the following solution of this
paradox.

An act of decay is an acausal phenomenon i.e. different particles
will decay at random instants with some probability density
$\varrho$ defined for a unit of proper time assigned to the
decaying particle (if the particle has no "memory" of its past
then $\varrho$ should be constant). The unit of $\varrho$ cannot
be expressed with the units of a mass and velocity only. Therefore
there must be a new fundamental constant having the unit of time,
or equivalently the unit of space. The new constant can have also
any other dimensionality that can be scaled into the unit of time
using mass and velocity. For the historical reason we assume this
fundamental constant to have the unit of an angular momentum - the
Planck's constant $\hslash$:
\begin{equation}
[\varrho] = \left[\frac{\mu c^2}{\hbar}\right].
\end{equation}
There is only one more fundamental constant known that has a
dimensionality allowing one to recover the unit of time - it is
the gravitational constant $G$, in a flat spacetime, however, it
cannot play any meaningful role.

Considering spontaneous acts of decay leads inevitably to a
preferred time-scale of the process. This scale, proportional to
$\hslash$ turns out to be, for the most processes, much shorter
than a typical time-scale of processes observed in the
macroscopical world. Therefore for the most of the
,,macroscopical'' processes the probabilities of possible decays
are approximately equal to one.

Describing the classical domain does not involve considering
systems containing a huge number of subsystems, but rather taking
into account the time-scales (or spatial scales) much larger than
the scales typical for the spontaneous processes. There are many
physical systems containing large number of particles, which
reveal quantum properties when observed in the proper scales. A
free neutron has an average lifetime of 10 minutes. This means
that a bomb triggered by a decay of a single neutron will explode
in a random moment, introducing a fundamental indeterminacy into
the macroscopical world. This example illustrates that it is not a
number of particles, but the typical time-scale that determines
the classical (or quantum) character of the process.

Going back to the example of exploding bomb we conclude that the
indeterminacy of the moment of explosion is still present,
although on a tiny time-scale. The probability of an explosion
within a microsecond is practically equal to unity and that is why
such an explosion may seem to be deterministic on the classical
scales.

Another interesting question arising from the fact that all the
decays must be spontaneous is the following: if we can't send
messages with sources of tachyons, how can we send messages with
sources of massive particles? The asymmetry originates from the
fact that we can shield a source of massive particles and modulate
the signal by uncovering the source, but we cannot do it with
sources of tachyons. From the diagrams in Fig.~\ref{Procesy2} it
follows that every object capable of absorbing tachyons must also
emit them, hence no shielding is possible.

\section{Superposition of world-lines\label{sec-superposition}}

\begin{figure}
\begin{center}
\epsfig{file=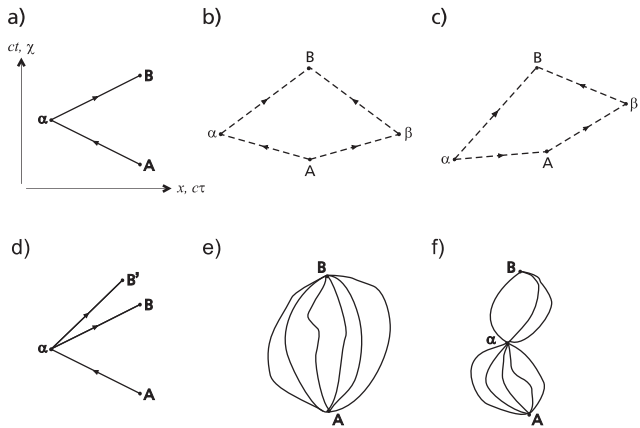} \caption{ \label{Young} \sf
\footnotesize Spacetime diagrams showing: a) motion of a particle
observed from two inertial frames. In the rest frame $(ct, x)$ a
tachyon departs from A, reflects in $\alpha$ and goes back to B.
In the frame moving infinitely fast $(c\tau,\chi)$ a source
$\alpha$ emits a particle which travels both towards A and B
simultaneously; b) particle emitted in a superposition state from
A, reflected at $\alpha$ and $\beta$ and detected at B; c) process
b) seen by a different inertial observer; d) particle emitted in
{\sf A} turns into a superposition after scattering in $\alpha$ -
another observer sees a triple superposition of a particle emitted
in $\alpha$; e) multiple non-intersecting paths allowed for a
particle moving between {\sf A} and {\sf B}; f) intersecting paths
- an example of a non-classical behavior of a particle that can be
described using the rules of the classical probability.}
\end{center}
\end{figure}

Consider the frame $(ct,x)$ in which a particle of a well-defined
momentum is emitted in {\sf A} - Fig.~\ref{Young}a). Let it be
reflected in $\alpha$ and arrive at {\sf B}. The particle that
reaches {\sf B} must first come across the path {\sf A}$\alpha$
and then across the path $\alpha${\sf B}. Therefore if the
observer places two detectors in the points intersecting the two
pathways then the detection of the particle on the path {\sf
A}$\alpha$ precludes the particle from being detected on the path
$\alpha${\sf B} and {\em vice versa}. If the particle is detected
on the path $\alpha${\sf B} then it could not have been absorbed
earlier on the path {\sf A}$\alpha$. From the Principle of
Relativity it follows that the same situation must take place in
all inertial frames. Another observer moving infinitely fast along
the $x$ axis, who describes the same spacetime with the
coordinates $(c\tau,\chi)$ will interpret the same course of
events in a different way. According to him there is a source
located at $\alpha$ that emits a particle with an uncertain
momentum. After the emission the particle can arrive either at the
point {\sf A} or {\sf B}, but if the observer places two detectors
on paths $\alpha${\sf A} and $\alpha${\sf B}, only one of these
detectors can absorb the particle emitted in $\alpha$. This
indicates that we have to attribute two world-lines to a single
particle, but when we try to localize the particle, its presence
is revealed on a single pathway only - we will call such a
phenomenon a {\em superposition} of world-lines.

Let us try to find a relativistically invariant expression
characterizing a spacetime path of a particle moving along two
world-lines. The unknown invariant expression ${\cal
P}_{\mbox{\scriptsize path}}$ for a given double path may depend
only on the relativistic invariants assigned to the space-time
path and the energy-momentum of the particle. There is {\em only
one} invariant not depending on the shape of the path - the
relativistic scalar product of the four-position and the
four-momentum - it will be called a phase $\phi$. For a particle
having the energy $E$, momentum $\boldsymbol{p}$ and moving along
a given pathway the phase equals:
\begin{equation}
\label{phase} \phi_{\mbox{\scriptsize path}} =
\hslash^{-1}\int_{\mbox{\scriptsize path}}(E\,\mbox{d}t -
{\boldsymbol p}\cdot \mbox{d}{\boldsymbol r}),
\end{equation}
where the proportionality constant has been introduced to keep the
phase dimensionless. The phase multiplied by the constant factor
$\hslash/mc^2$ can be also interpreted as the proper time or a
classical action associated with the path. Let us investigate how
such a double path transforms to another frame of reference.
Consider a situation when a tachyonic particle is emitted in {\sf
A} - Fig.~\ref{Young}b) and reflected in $\alpha$ and $\beta$ so
that speed is decreased on both paths. None of the two paths is
distinguished therefore the invariant ${\cal P}$ should be a
symmetric function of the phases calculated for the two paths:
\begin{equation}
\label{symmetricprobability} {\cal P}(\phi_1,\phi_2) = {\cal P}
(\phi_2,\phi_1),
\end{equation}
where the indices refer to the paths {\sf A}$\alpha${\sf B} and
{\sf A}$\beta${\sf B} traversed by the particle -
Fig.~\ref{Young}b). Observing the same process from a moving frame
of reference gives a different picture of the situation -
Fig.~\ref{Young}c). The moving observer claims that the particle
is emitted in $\alpha$ and follows two paths. One of them leads
directly to {\sf B} and on the other the particle is scattered
twice - in {\sf A} and $\beta$, and consecutively reaches {\sf B}.
In this inertial frame the invariant ${\cal P}$ is described by
different paths $1'$ referring to $\alpha${\sf B} and $2'$
referring to $\alpha${\sf A}$\beta${\sf B} with the respective
phases:
\begin{eqnarray}
\phi_{1'} &=& \phi_1-\phi_{{\sf A}\alpha}, \nonumber \\
\phi_{2'} &=& \phi_2+\phi_{\alpha{\sf A}},
\end{eqnarray}
where $\phi_{\alpha{\sf A}} = -\phi_{{\sf A}\alpha}$. For an
arbitrary process described by a closed space-time loop, as in
Fig.~\ref{Young}b) or \ref{Young}c) the phase $\phi_{\alpha{\sf
A}}$ can take an arbitrary value, therefore from the condition
\begin{equation}
\label{symmetricprobability2} {\cal P}(\phi_1,\phi_2) = {\cal P}
(\phi_{1'},\phi_{2'}),
\end{equation}
and the equation \eqref{symmetricprobability} follows that ${\cal
P}$ must be a symmetric function of the phase difference only
${\cal P}(|\phi_1-\phi_2|)$. We see that such an invariant cannot
be factorized into a sum of functions $P$ depending on the single
paths only:
\begin{equation}
\label{nonseparability} {\cal P}(|\phi_1-\phi_2|)\neq P(\phi_1) +
P(\phi_2).
\end{equation}

The problem of the particle's motion along two spacetime paths can
be generalized to multiple paths using the induction method.
Suppose a particle emitted in {\sf A} and reflected in $\alpha$
finds itself in a superposition of two world lines -
Fig.~\ref{Young}d). One of the lines is directed towards the event
{\sf B}, while the other one towards some other event {\sf B'}.
Another observer viewing the process finds the particle in a
superposition of three world-lines originating from the event
$\alpha$. The further generalization is straightforward.

A relativistic invariant describing $n$ non-intersecting spacetime
paths linking two events will be denoted ${\cal P}^{(n)} (\phi_1,
\phi_2, \ldots,\phi_n)$ - see Fig.~\ref{Young}e). In order to
determine its explicit form we will postulate the following four
axioms. We will assume that the invariant ${\cal P}$ must be a
smooth function of phases only and does not depend on the paths'
topology. The function must also be completely symmetric, i.e. for
an arbitrary permutation $\pi$ of an $n$-element set we have:
\begin{equation}
\label{axiom-symmetry} {\cal P}^{(n)} (\phi_1, \phi_2,
\ldots,\phi_n) = {\cal P}^{(n)} (\phi_{\pi(1)}, \phi_{\pi(2)},
\ldots,\phi_{\pi(n)}).
\end{equation}
The third axiom demands that the function does not depend on the
arrow of time, therefore it must be invariant under the inversion:
\begin{equation}
\label{axiom-inverse} {\cal P}^{(n)} (\phi_1, \phi_2,
\ldots,\phi_n) = {\cal P}^{(n)} (-\phi_1, -\phi_2,
\ldots,-\phi_n).
\end{equation}
In order to introduce the last axiom let us go back to the
discussion of the expression \eqref{nonseparability}. According to
this equation in the simplest case of the two paths the invariant
${\cal P}$ does not factorize into a sum of two expressions, as
required for the classical probability \cite{Cox1946}. This is the
consequence of a non-classical character of a superposition.
However there is a special case when the rules of the classical
probability may apply to the presently developed formalism. The
fourth axiom expresses the probability-like character of the
invariant ${\cal P}$. Consider a set of intersecting paths shown
in Fig.~\ref{Young}f) - if $n$ paths linking {\sf A} and $\alpha$
traversed by a particle intersect with $m$ paths between $\alpha$
and {\sf B} then the presence of a particle in a spacetime
location $\alpha$ is certain. In this case we can apply the law of
composition of classical probabilities. If our invariant function
${\cal P}$ is to express the probability for a particle to take a
given composite path then in the considered case the probability
should be a product of two probabilities for the motion along the
paths linking {\sf A} with $\alpha$ and the paths linking $\alpha$
with {\sf B}. This is the content of our last axiom:
\begin{widetext}
\begin{equation}
\label{axiom-probability} {\cal P}^{(n)} (\phi_1, \phi_2,
\ldots,\phi_n){\cal P}^{(m)} (\xi_1, \xi_2, \ldots,\xi_m)
 = {\cal P}^{(nm)}(\phi_1+\xi_1, \phi_1+\xi_2, \phi_1+\xi_3, \ldots, \phi_n+\xi_m).
\end{equation}
Since we can permutate the arguments appearing on the left-hand
side of the equation, the arguments of the function on the
right-hand side must involve sums of all the possible pairs of
phases $\phi_i+\xi_j$. In the above condition we have also used
the first axiom assuming that the invariant expression describing
$n$ non-intersecting paths depicted in Fig.~\ref{Young}e)
coincides with the expression describing $n$ intersecting paths
shown in Fig.~\ref{Young}f). Let us underline that the above set
of axioms is a set of necessary, but not sufficient conditions for
the invariant ${\cal P}^{(n)} (\phi_1, \phi_2, \ldots,\phi_n)$ to
define a probability.

One can easily check that the following function is smooth and
obeys the conditions \eqref{axiom-symmetry},
\eqref{axiom-inverse}, and \eqref{axiom-probability}:
\begin{equation}
\label{AEfunction} {\cal P}^{(n)}(\phi_1, \phi_2, \ldots, \phi_n)
= \frac{1}{n^{\mbox{\scriptsize \em \c{A}}}} \left(
e^{\alpha\phi_1} + e^{\alpha\phi_2} + \ldots +
e^{\alpha\phi_n}\right) \left( e^{-\alpha\phi_1} +
e^{-\alpha\phi_2} + \ldots + e^{-\alpha\phi_n}\right),
\end{equation}
\end{widetext}
where {\em \c{A}} and $\alpha$ are arbitrary constants. In
Appendix \ref{proof-wavefunction} we show that a general solution
of the problem is given by a product of arbitrarily many basic
solutions of the form \eqref{AEfunction}.

For an infinite number of paths the expression \eqref{AEfunction}
becomes infinite for any real $\alpha$. In order to keep the
invariant finite for arbitrary phases one has to take into account
only imaginary $\alpha=\pm i|\alpha|$. The modulus $|\alpha|$ can
be associated with an arbitrary value of the Planck's constant
$\hslash$, so without the loss of generality we can assume
$|\alpha|=1$ for a basic solution. If we consider $n$ identical
paths and demand ${\cal P}^{(n)}(\phi, \phi, \ldots, \phi) = {\cal
P}^{(1)}(\phi)$ we obtain the condition $\mbox{\em \c{A}}=2$.
Hence we can introduce the following notation:
\begin{equation}
\langle B|A\rangle = \frac{1}{n}\left(e^{i\phi_1} + e^{i\phi_2} +
\ldots + e^{i\phi_n}\right),
\end{equation}
where $A$ and $B$ are two spacetime events and the sum extends
over all $n$ allowable paths connecting events $A$ and $B$. In
this notation we have $\langle A|B\rangle^* = \langle B|A\rangle$
and our simplest probability-like relativistic invariant reduces
to:
\begin{equation}
{\cal P}^{(n)}(\phi_1, \phi_2, \ldots, \phi_n)=\langle
A|B\rangle\langle B|A\rangle.
\end{equation}
So it goes. Considering infinite number of paths linking two
spacetime events we end up with the Feynmanian theory in which one
has to take into account a sum over all possible histories with
complex amplitudes based on classical action attributed to each
path.

This picture can be intuitively understood on the ground of a
weird hypothesis that in the class of superluminal inertial frames
there are three temporal dimensions and each of them flows exactly
as it happens with the single temporal dimension in subluminal
frames. The last statement demands the abandonment of the concept
of the world-{\em line} when considering superluminal observers.
For such observers no arrow of time is preferred therefore it is
natural to assume that all the objects observed by a superluminal
observer grow older along all directions of time
$\boldsymbol{\tau}$. Consequently in the arbitrary superluminal
frame every physical object traversing a given point in the
spacetime should have a three-dimensional world-line - a
world-{\em sphere} attributed to it. The above peculiar principle
leads to the well-known experimental facts observed in the
subluminal frames. From the Principle of Relativity it follows
that in the arbitrary subluminal frame every particle must also
have a space-time {\em sphere} attributed to it in each space-time
location of the particle. The last statement is known as a part of
the Huygens' Principle originally formulated to describe light and
many years later discovered to apply also to any matter. The
Principle states that {\em every point in space traversed by light
is a source of a new spherical wave}. What follows is that in
order to describe a motion of a particle one has to take into
account all possible space-time paths, which we have just
concluded on the ground of the four elementary axioms.

\section{Symmetries\label{sec-CPT}} Let us consider a scenario
when a particle described by one of the four-momentum
\eqref{energymomentum} exists in Nature. In order to calculate its
energy and momentum one has to determine not only the particle's
mass and velocity, but also an additional parameter $\mbox{\em
\'{s}}$. The only known scalar intrinsic degree of freedom of a
free, uncharged particle is its helicity. Let us therefore study
the case, when $\mbox{\em \'{s}}$ has the symmetry properties of
the helicity. The time reversal transformation {\sf T} leaves the
helicity unchanged, while the spatial reflection {\sf P} changes
its sign:
\begin{eqnarray}
{\sf T}\mbox{\em \'{s}} &=& \mbox{\em \'{s}}\nonumber \\
{\sf P}\mbox{\em \'{s}} &=& -\mbox{\em \'{s}}.
\end{eqnarray}
Suppose that a process of decay of a massive particle and a
particle described by \eqref{energymomentum} takes place, as
depicted in Fig.~\ref{Procesy2}a). We assume that the total energy
and momentum is conserved. The time reversal operation ${\sf T}$
changes signs of velocities, therefore both four-vectors
\eqref{energymomentum} transform identically and ${\sf T}$ is a
symmetry of the process. However, under the parity transformation
${\sf P}$ the considered four-vectors change in a different
fashion, which means that after the spatial reversal ${\sf P}$
neither energy, nor momentum will be conserved in the process.
This shows that the process will have no right to take place.

These considerations are based on the assumption that $\mbox{\em
\'{s}}$ has the properties of the helicity. One can, however, take
into account particles of different types, characterized by
$\mbox{\em \'{s}}$ obeying other transformation rules, so that
other symmetries apply to the considered types od decays, in
particular the time reversal wouldn't have to be a symmetry. If we
assume that the conjugation ${\sf C}$ reverses the direction of
$\mbox{\em \'{s}}$ (as discussed in detail in Appendix
\ref{proof-fourmomentum}) and we demand the overall operation of
${\sf CPT}$ to be a symmetry then either the parity or the time
reversal symmetry must be broken in the interaction of given
particles. The latter is represented by:
\begin{eqnarray}
\label{angularmomentumsymmetries}
{\sf T}\mbox{\em \'{s}} &=& -\mbox{\em \'{s}}\nonumber \\
{\sf P}\mbox{\em \'{s}} &=& \mbox{\em \'{s}}.
\end{eqnarray}
Existence of tachyons interacting with matter, whose energy and
momentum depend on the velocity and the helicity parameter
$\mbox{\em \'{s}}$ according to the expression
\eqref{energymomentum} leads to the violation of the parity
symmetry, which, as we know is not obeyed in the weak
interactions. This suggests that tachyonic particles play some
role in the weak interactions and the present mode of description
of these interactions should be understood only as an effective
theory.

\section{Conclusions\label{Conclusions}}
We have shown that the disturbances of causal laws resulting from
the extension of the Principle of Relativity to superluminal
frames coincide with the laws known from the basic postulates of
Quantum Theory. There's a method in the madness - it follows that
Quantum Theory is relativistic to the roots and the term
"non-relativistic quantum mechanics" is an oxymoron like
"non-relativistic electrodynamics". The presented results do not
indicate that the tachyons must exist, however it would be
surprising if they didn't. There are not too many new predictions,
except for the suspicion that the tachyons should take part in the
weak interactions. Moreover the deeper understanding of the roots
of Quantum Theory may be helpful in constructing the still unknown
quantum theory of gravity. It seems necessary that such a theory
should take into account not only subluminal, but also
superluminal class of local observers.

\begin{acknowledgements}
I would like to thank Iwo Bia{\l}ynicki-Birula, Szymon
Charzy\'{n}ski, Tomasz Mas{\l}owski, Bogdan Mielnik, Jan
Mostowski, Krzysztof Pachucki, and Pawe{\l} Zi\'{n} for
interesting conversations.
\end{acknowledgements}

\begin{appendix}

\section{Derivation of the energy-momentum four-vectors\label{proof-fourmomentum}}

A four-vector $A^\square \equiv (A^0,\boldsymbol{A})$, by the
definition transforms to the inertial frame moving with the
velocity $\boldsymbol{V}$ according to the
formulas:
\begin{equation} \label{3d-fourvectortransform}
\begin{split}
{A^0}' & = \frac{A^0-\frac{{\boldsymbol A}\cdot{\boldsymbol
V}}{c}}{\sqrt{1-\frac{V^2}{c^2}}}, \\
\boldsymbol{A}' & = \boldsymbol{A}-\frac{{\boldsymbol
A}\cdot{\boldsymbol V}}{V^2}{\boldsymbol V} +
\frac{\frac{{\boldsymbol A}\cdot{\boldsymbol V}}{V^2}{\boldsymbol
V}-A^0\frac{\boldsymbol V}{c}} {\sqrt{1-\frac{V^2}{c^2}}}.
\end{split}
\end{equation}
We are looking for all the four-vectors $A^\square$ that do
transform in a covariant way, thus obeying the equation:
\begin{equation}
\label{3d-fourvectortransform2} {A^\square}'(\boldsymbol{v},{\cal
A}^\square) = {A^\square}(\boldsymbol{v}',{\cal A}'^\square),
\end{equation}
where $\boldsymbol{v}$ is velocity transforming according to the
formula:
\begin{equation}
\label{velocitytransform} \boldsymbol{v}' =
\frac{\sqrt{1-\frac{V^2}{c^2}}\left(\boldsymbol{v}-\frac{{\boldsymbol
v}\cdot{\boldsymbol V}}{V^2}{\boldsymbol V}\right)-\boldsymbol
V+\frac{{\boldsymbol v}\cdot{\boldsymbol V}}{V^2}{\boldsymbol V}}
{1-\frac{\boldsymbol{v}\cdot\boldsymbol{V}}{c^2}},
\end{equation}
and ${\cal A}^\square \equiv ({\cal A}^0,\boldsymbol{\cal A})$ is
an additional parameter - a value of the four-vector $A^\square$
in a selected inertial frame.

It turns out that there are only four linearly independent
four-vectors $A^\square$ obeying the condition
\eqref{3d-fourvectortransform2}:
\begin{subequations}
\label{3d-allfourvectors}
\begin{eqnarray}
\label{3d-realmass}&&\left(\frac{1}{\sqrt{1-v^2/c^2}},
\frac{\boldsymbol{v}/c}{\sqrt{1-v^2/c^2}}\right),\\
\label{3d-newmass}&&\left(\frac{{\boldsymbol s}\cdot{\boldsymbol
v}/c} {\sqrt{1-v^2/c^2}},
\boldsymbol{s}-\frac{\boldsymbol{s}\cdot\boldsymbol{v}}{v^2}\boldsymbol{v}+\frac{(\boldsymbol{s}\cdot\boldsymbol{v})\boldsymbol{v}/v^2}{\sqrt{1-v^2/c^2}}\right),\\
\label{3d-tachionmass}&&\left(
\frac{\mbox{sgn}\left(\boldsymbol{s}\cdot\boldsymbol{w}\right)}{\sqrt{w^2/c^2-1}},
\frac{\mbox{sgn}\left(\boldsymbol{s}\cdot\boldsymbol{w}\right)\boldsymbol{w}/c}{\sqrt{w^2/c^2-1}}\right),\\
\label{3d-newtachion}&&\left(
\frac{\frac{w^2/c^2}{\sqrt{w^2/c^2-1}}-\left|\boldsymbol{s}\cdot\boldsymbol{w}\right|/c
} {\left|\boldsymbol{s}\cdot\boldsymbol{w}\right|/c
-\sqrt{w^2/c^2-1}},
\frac{\frac{\boldsymbol{w}/c}{\sqrt{w^2/c^2-1}}-\mbox{sgn}(\boldsymbol{s}\cdot\boldsymbol{w})\boldsymbol{s}
} {\left|\boldsymbol{s}\cdot\boldsymbol{w}\right|/c
-\sqrt{w^2/c^2-1}} \right),\nonumber \\
\end{eqnarray}
\end{subequations}
where the function $\mbox{sgn}(x)$ returns the sign of its
argument $x$ and $\boldsymbol{s}$ is a dimensionless unit vector
or pseudo-vector undergoing a Wigner-Thomas precession by Lorentz
transform. The first pair of four-vectors is defined for
subluminal velocities $|\boldsymbol{v}|<c$ and the second pair for
the superluminal velocities $|\boldsymbol{w}|>c$. Moreover the
parameters determining the four-vectors \eqref{3d-tachionmass} and
\eqref{3d-newtachion} must obey the condition
$w^2-c^2<(\boldsymbol{s}\cdot\boldsymbol{w})^2$. The proof is
following.

Suppose that the frame of reference for which $A^\square = {\cal
A}^\square$ is the frame for which $\boldsymbol{v}=0$ then the
transition to a frame moving with a relative velocity
$-\boldsymbol{V}$, for which
$\boldsymbol{v}'=\boldsymbol{\boldsymbol{V}}$ yields:
\begin{eqnarray}
\label{3d-fourvectortransform3} A^0(\boldsymbol{V},{\cal
A}^\square) &=& \frac{{\cal A}^0+\frac{{\boldsymbol {\cal A
}}\cdot{\boldsymbol V}}{c}}
{\sqrt{1-\frac{V^2}{c^2}}} ,\nonumber \\
\boldsymbol{A}(\boldsymbol{V},{\cal A}^\square) &=&
\boldsymbol{\cal A}-\frac{{\boldsymbol {\cal A}}\cdot{\boldsymbol
V}}{V^2}{\boldsymbol V}+\frac{\frac{{\boldsymbol {\cal
A}}\cdot{\boldsymbol V}}{V^2}{\boldsymbol V}+{\cal A}^0
\frac{\boldsymbol V}{c}} {\sqrt{1-\frac{V^2}{c^2}}}, \nonumber \\
\end{eqnarray}
Assuming that ${\cal A}^\square \equiv (1,0)$ and replacing
$\boldsymbol{V}$ with $\boldsymbol{v}$ we obtain the inside-cone
four-vector \eqref{3d-realmass}. Taking ${\cal A}^\square \equiv
(0,\boldsymbol{s})$ we get the outside-cone four-vector
\eqref{3d-newmass}.

Let us discuss the transformation rules for the direction
$\boldsymbol{s}$ parameterizing the four-vector
\eqref{3d-newmass}. Let us denote the Lorentz transformation for
the velocity $\boldsymbol{V}$ with $\Lambda(\boldsymbol{V})$ and
the velocity transformation \eqref{velocitytransform} with
$\Gamma(\boldsymbol{V})$. The covariance condition
\eqref{3d-fourvectortransform2} in an arbitrary inertial frame
takes the following form:
\begin{equation}
\label{3d-fourvectortransform5}
\Lambda(\boldsymbol{V})A^\square(\boldsymbol{v},\boldsymbol{s}) =
A^\square(\Gamma(\boldsymbol{V})\boldsymbol{v},\boldsymbol{s}'),
\end{equation}
where $\boldsymbol{s}'$ is unknown. Using the definition of
$\boldsymbol{s}$: $A^\square(\boldsymbol{v},\boldsymbol{s}) =
\Lambda(-\boldsymbol{v})(0,\boldsymbol{s})$ and the property of
boosts $\Lambda^{-1}(\boldsymbol{V}) = \Lambda(-\boldsymbol{V})$
we obtain the relation:
\begin{equation}
(0, \boldsymbol{s}') =
\Lambda(\Gamma(\boldsymbol{V})\boldsymbol{v})
\Lambda(\boldsymbol{V}) \Lambda(-\boldsymbol{v})(0,
\boldsymbol{s}).
\end{equation}
The above series of boosts relating three inertial frames in a
non-collinear relative motion is a spatial rotation
\cite{Wigner1939} called the Wigner-Thomas rotation. Such
transformation does not affect the temporal coordinate of the
four-vector and therefore there are no further complications in
the transformation law of the four-vector \eqref{3d-realmass}.

Let us now consider a situation, when in a given frame of
reference the velocity parameter has the direction
$\boldsymbol{s}$ and an infinite magnitude. Now we choose this
frame to define ${\cal A}^\square$. The transformation formula
\eqref{velocitytransform} with the frame's relative velocity
$-\boldsymbol{V}$ after replacing $\boldsymbol{v}'$ with
$\boldsymbol{w}$ yields:
\begin{equation}
\label{velocitytransform2}
\frac{\boldsymbol{s}\cdot\boldsymbol{V}}{c^2}\boldsymbol{w}
=\sqrt{1-\frac{V^2}{c^2}}\left(\boldsymbol{s}-\frac{\boldsymbol{s}\cdot\boldsymbol{V}}{V^2}\boldsymbol{V}\right)+
\frac{\boldsymbol{s}\cdot\boldsymbol{V}}{V^2}\boldsymbol{V}.
\end{equation}
Let us notice that reversing the sign of $\boldsymbol{s}$ in the
equation \eqref{velocitytransform2} does not change the equation
itself, therefore the relations between the velocities
$\boldsymbol{w}$ and $\boldsymbol{V}$ remain unchanged. This means
that $\boldsymbol{s}$ can have transformation properties of a
vector or a pseudo-vector, which has very important implications
to the symmetries of the collision processes discussed in
Sec.~\ref{sec-CPT}.

The transformation law for the four-vector
\eqref{3d-fourvectortransform2} in the considered frame of
reference has the form:
\begin{eqnarray}
\label{3d-fourvectortransform4} A^0(\boldsymbol{w},{\cal
A}^\square) &=& \frac{{\cal A}^0+\frac{{\boldsymbol {\cal A}}\cdot
{\boldsymbol
V}}{c}}{\sqrt{1-\frac{V^2}{c^2}}},\nonumber \\
\boldsymbol{A}(\boldsymbol{w},{\cal A}^\square) &=&
\boldsymbol{{\cal A}}-\frac{{\boldsymbol {\cal
A}}\cdot{\boldsymbol V}}{V^2}{\boldsymbol
V}+\frac{\frac{{\boldsymbol {\cal A}}\cdot{\boldsymbol V}}
{V^2}{\boldsymbol V}+{\cal A}^0\frac{\boldsymbol V}{c}}
{\sqrt{1-\frac{V^2}{c^2}}}. \nonumber \\
\end{eqnarray}
Taking ${\cal A}^\square\equiv(0,\boldsymbol{s})$
($\boldsymbol{s}$ is the only preferred direction in space;
moreover this condition guarantees that $\boldsymbol{A}$ being a
candidate for momentum has the direction of velocity) and using
\eqref{velocitytransform2} we get:
\begin{eqnarray}
\label{fourvector-tobedone} A^0(\boldsymbol{w},\boldsymbol{s}) &=&
\frac{\frac{{\boldsymbol s}\cdot
{\boldsymbol V}}{c}}{\sqrt{1-\frac{V^2}{c^2}}},\nonumber \\
\boldsymbol{A}(\boldsymbol{w},\boldsymbol{s}) &=&
\frac{\frac{{\boldsymbol s}\cdot {\boldsymbol
V}}{c}\frac{\boldsymbol{w}}{c}} {\sqrt{1-\frac{V^2}{c^2}}}.
\end{eqnarray}
The above four-vector is expressed as a function of the velocity
$\boldsymbol{V}$ that can be interpreted as the relative velocity
of a frame in which $\boldsymbol{w}$ attains infinite magnitude
and the direction $\boldsymbol{s}$. We wish now to express the
four-vector as an explicit function of $\boldsymbol{w}$ and
$\boldsymbol{s}$.

Let us take a scalar product of the equation
\eqref{velocitytransform2} with the (pseudo)vector
$\boldsymbol{s}$. We obtain the condition
$(\boldsymbol{s}\cdot\boldsymbol{V}) (\boldsymbol{s}\cdot
\boldsymbol{w})>0$. Taking a square of the equation
\eqref{velocitytransform2} and using the above identity we get:
\begin{equation}
\label{projection} \frac{{\boldsymbol s}\cdot {\boldsymbol V}}{c}
=\frac{\sqrt{1-\frac{V^2}{c^2}}}{\sqrt{\frac{w^2}{c^2}-1}}\,
\mbox{sgn}(\boldsymbol{s}\cdot \boldsymbol{w}).
\end{equation}
After putting this expression into \eqref{fourvector-tobedone} we
obtain the outside-cone four-vector \eqref{3d-tachionmass}.

The last of the four-vectors \eqref{3d-allfourvectors} is obtained
by assuming in equations \eqref{3d-fourvectortransform4} the
condition ${\cal A}=(1, 0)$ leading to:
\begin{eqnarray}
\label{fourvector-tobedone2} A^0(\boldsymbol{w},\boldsymbol{s})
&=&
\frac{1}{\sqrt{1-\frac{V^2}{c^2}}},\nonumber \\
\boldsymbol{A}(\boldsymbol{w},\boldsymbol{s}) &=&
\frac{\frac{\boldsymbol{V}}{c}}{\sqrt{1-\frac{V^2}{c^2}}}.
\end{eqnarray}
To express the above formulas with the velocity $\boldsymbol{w}$
and $\boldsymbol{s}$ we will transform the equation
\eqref{velocitytransform2} to a new form. Using the formula
\eqref{projection} we get:
\begin{equation}
\label{velocityreversed}
\frac{\boldsymbol{V}}{1+\sqrt{1-\frac{V^2}{c^2}}} = \boldsymbol{w}
- \mbox{sgn}\left(\boldsymbol{s}\cdot\boldsymbol{w}\right)
\sqrt{w^2-c^2}\boldsymbol{s}.
\end{equation}
Taking a square of the equation \eqref{velocityreversed} we
determine the Lorentz factor:
\begin{equation}
\sqrt{1-\frac{V^2}{c^2}} =
\frac{\left|\boldsymbol{s}\cdot\boldsymbol{w}\right|-\sqrt{w^2-c^2}}
{\frac{w^2}{\sqrt{w^2-c^2}}-\left|\boldsymbol{s}\cdot\boldsymbol{w}\right|}.
\end{equation}
Hence the explicit form of the relative velocity of the two
considered inertial frames:
\begin{equation}
\label{velocityreversed2} \boldsymbol{V} = c^2\frac{\boldsymbol{w}
- \mbox{sgn}\left(\boldsymbol{s}\cdot\boldsymbol{w}\right)
\sqrt{w^2-c^2}\boldsymbol{s}}{w^2 -
|\boldsymbol{s}\cdot\boldsymbol{w}| \sqrt{w^2-c^2}},
\end{equation}
and the inside-cone four-vector \eqref{3d-newtachion}. Taking a
scalar product of the equation \eqref{velocityreversed2} with
$\boldsymbol{w}$ we obtain the equality
$\boldsymbol{w}\cdot\boldsymbol{V}=c^2$ determining the relation
between the superluminal velocity $\boldsymbol{w}$ and the
velocity $\boldsymbol{V}$ of the inertial frame in which
$\boldsymbol{w}$ becomes infinite.

The covariance condition \eqref{3d-fourvectortransform5} for the
four-vector \eqref{3d-tachionmass} in an arbitrary inertial frame
leads to the equation:
\begin{equation}
\frac{\mbox{sgn}\left(\boldsymbol{s}\cdot\boldsymbol{w}\right)}{\sqrt{w^2/c^2-1}}
\frac{1-\frac{\boldsymbol{w}\cdot\widetilde{\boldsymbol{V}}}{c^2}}{\sqrt{1-\widetilde{V}^2/c^2}}=
\frac{\mbox{sgn}\left(\boldsymbol{s}'\cdot\boldsymbol{w}'\right)}{\sqrt{w'^2/c^2-1}},
\end{equation}
where $\boldsymbol{w}'$ is the velocity and $\boldsymbol{s}'$ the
direction parameterizing the four-vector \eqref{3d-tachionmass} in
a new inertial frame moving with a relative velocity
$\widetilde{\boldsymbol{V}}$. Taking the square of the equation
\eqref{velocitytransform} and using it in the above expression we
obtain:
\begin{equation}
\label{signum-transform}
\mbox{sgn}\left(\boldsymbol{s}'\cdot\boldsymbol{w}'\right) =
\mbox{sgn}\left(\boldsymbol{s}\cdot\boldsymbol{w}\right)
\mbox{sgn}\left(c^2-\boldsymbol{w}\cdot\widetilde{\boldsymbol{V}}\right).
\end{equation}
During the transformation to a new inertial frame, the direction
$\boldsymbol{s}$ follows in general the Wigner-Thomas precession.
That's why the sign of the energy and momentum of a tachyon is
changed if and only if the relative velocity
$\widetilde{\boldsymbol{V}}$ of a new inertial frame is such that
$\boldsymbol{w}\cdot\widetilde{\boldsymbol{V}}>c^2$, i.e. the
tachyon becomes an anti-tachyon. Let us find the transformation
law for the direction $\boldsymbol{s}$. From the covariance
requirement \eqref{3d-fourvectortransform5} we obtain:
\begin{equation}
\Lambda(\widetilde{\boldsymbol{V}})A^\square(\boldsymbol{w},\boldsymbol{s})
= A^\square(\boldsymbol{w}',\boldsymbol{s}'),
\end{equation}
where $\widetilde{\boldsymbol{V}}$ is the velocity of the new
inertial frame,
$\boldsymbol{w}'=\Gamma(\widetilde{\boldsymbol{V}})\boldsymbol{w}$
and $\boldsymbol{s}'$ is unknown. This condition for the
four-vectors \eqref{3d-tachionmass} and \eqref{3d-newtachion}
yields, respectively:
\begin{eqnarray}
\label{s-tachiontransform}
\Lambda(\boldsymbol{V}(\boldsymbol{w}',\boldsymbol{s}'))
\Lambda(\widetilde{\boldsymbol{V}})
\Lambda(-\boldsymbol{V}(\boldsymbol{w},\boldsymbol{s}))(0,\boldsymbol{s})
&=& (0,\boldsymbol{s}'),\nonumber \\
\Lambda(\boldsymbol{V}(\boldsymbol{w}',\boldsymbol{s}'))
\Lambda(\widetilde{\boldsymbol{V}})
\Lambda(-\boldsymbol{V}(\boldsymbol{w},\boldsymbol{s}))(1,0)
&=& (1,0),\nonumber \\
\end{eqnarray}
where $\boldsymbol{V}(\boldsymbol{w},\boldsymbol{s})$ is given by
the expression \eqref{velocityreversed2}. The above equations can
be satisfied only if the three consecutive Lorentz transformations
on the left-hand side are equivalent to some Wigner-Thomas
rotation. This is possible if and only if transformations'
arguments are related via the velocity transformation
\eqref{velocitytransform}:
\begin{equation}
\boldsymbol{V}(\boldsymbol{w}',\boldsymbol{s}') =
\boldsymbol{V}'(\boldsymbol{w},\boldsymbol{s}),
\end{equation}
where $\boldsymbol{V}'
=\Gamma(\widetilde{\boldsymbol{V}})\boldsymbol{V}$. Substituting
it into the first of the equations \eqref{s-tachiontransform} we
obtain the condition defining the parameter $\boldsymbol{s}$ in a
frame moving with the velocity $\widetilde{\boldsymbol{V}}$:
\begin{equation}
\boldsymbol{s}' =
\Lambda(\Gamma(\widetilde{\boldsymbol{V}})\boldsymbol{V}(\boldsymbol{w},\boldsymbol{s}))
\Lambda(\widetilde{\boldsymbol{V}})
\Lambda(-\boldsymbol{V}(\boldsymbol{w},\boldsymbol{s}))\boldsymbol{s}.
\end{equation}
At the end, let us notice that the magnitude of the velocity
$\boldsymbol{V}(\boldsymbol{w},\boldsymbol{s})$ can't exceed the
magnitude of $c$. Taking the square of the equation
\eqref{velocityreversed} and imposing this condition we obtain the
following inequality:
\begin{equation}
w^2-c^2<(\boldsymbol{s}\cdot\boldsymbol{w})^2,
\end{equation}
that limits the choice of possible parameters of the four-vectors
\eqref{3d-tachionmass} and \eqref{3d-newtachion} in subluminal
frames.

Energy and momentum of a tachyon with a mass parameter $\mu$,
velocity $\boldsymbol{w}$ and "helicity" $\mbox{\em \'{s}}=
\mbox{sgn}(\boldsymbol{s}\cdot\boldsymbol{w})$ given by the
expression \eqref{3d-tachionmass} or \eqref{energymomentum} have
the properties that energy tends to zero and momentum decreases to
the minimum value $\mu c$ when the velocity increases. Energy and
momentum increases to infinity when the velocity tends to the
velocity of light, so crossing the border of $|\boldsymbol{w}|=c$
is not energetically possible. Therefore in the two-dimensional
case the behavior of tachyons is fully analogical to the behavior
if massive particles if only we interchange the temporal and
spatial components of the considered four-vectors.

From the velocity transformation formula \eqref{velocitytransform}
for the superluminal velocities one can conclude that observing a
tachyon moving with the velocity $\boldsymbol{w}$ from a reference
frame following the tachyon with a velocity $\boldsymbol{V}$
increases, not decreases the tachyon's velocity. When the velocity
$\boldsymbol{V}$ of the inertial frame is such that
$\boldsymbol{w}\cdot\boldsymbol{V}=c^2$, the tachyon escapes with
an infinite velocity. In an inertial frame such that
$\boldsymbol{w}\cdot\boldsymbol{V}>c^2$, the tachyon's energy
becomes negative and its momentum gets reversed in respect to the
tachyon's velocity. In the spirit of Feynman one can say that in
this inertial frame the tachyon becomes its anti-particle
\cite{Feynman1949}. If we accompany each world-line with an arrow
pointing towards the direction of the propagation in spacetime
then a tachyon that moves in a stationary frame with the velocity
$\boldsymbol{w}$ ahead in time, observed from the inertial frame
for which $\boldsymbol{w}\cdot\boldsymbol{V}>c^2$ moves backwards
in time. To make sure that the emission of a tachyon is fully
equivalent to an absorption of an anti-tachyon we have to prove
that the energy and momentum reverse their signs in the same
inertial frame in which the velocity becomes infinite, so that
reversing the sign of $\boldsymbol{s}$ is equivalent to changing a
tachyon into its anti-particle. We have shown that it happens
indeed - the sign function that regulates the sign of energy and
momentum in expression \eqref{3d-tachionmass} obeys the
transformation rule \eqref{signum-transform}. Therefore one can
always reinterpret the emitted anti-tachyon with negative energy
as an absorbed tachyon with positive energy. The interchange
procedure is equivalent to changing the sign of $\boldsymbol{s}$
and must be related to the charge conjugation operation {\sf C},
as discussed in Sec.~\ref{sec-CPT}.

\section{Derivation of all the probability-like relativistic invariants\label{proof-wavefunction}}

Let ${\cal P}^{(n)}(\phi_1, \phi_2, \ldots, \phi_n)$ and ${\cal
R}^{(n)}(\phi_1, \phi_2, \ldots, \phi_n)$ be arbitrary smooth
functions obeying all the conditions \eqref{axiom-symmetry},
\eqref{axiom-inverse}, and \eqref{axiom-probability}. We find that
the product ${\cal P}^{(n)}(\phi_1, \phi_2, \ldots, \phi_n) {\cal
R}^{(n)}(\phi_1, \phi_2, \ldots, \phi_n)$ is also smooth and obeys
the above axioms. Therefore in order to obtain a general solution
obeying all the axioms, we need to find all the special solutions
that are irreducible to the product of other solutions. Consider a
Taylor expansion of a smooth, completely symmetric function ${\cal
P}^{(n)}(\phi_1, \phi_2, \ldots, \phi_n)$. From the Cauchy's
theorem on symmetric many-variable polynomials
\cite{Sierpinski1946} it follows that it can be expressed in terms
of a power series of the symmetric functions $E^{(k)}(\phi_1,
\phi_2, \ldots, \phi_n)=\sum_{i=1}^n \phi_i^k$ in the form:
\begin{equation}
\label{expansion}
\begin{split}
{\cal P}^{(n)}(\phi_1, \phi_2, \ldots, \phi_n) = \sum_{l=0}^\infty
\sum_{k_1,k_2,\ldots,k_{l}=1}^\infty
\alpha^{(n)}_{k_1,k_2,\ldots,k_l} \\
\times E^{(k_1)}(\phi_1,\phi_2,\ldots,\phi_n)\cdots
E^{(k_l)}(\phi_1,\phi_2,\ldots,\phi_n).
\end{split}
\end{equation}
The set of symmetric polynomials $E^{(k)}(\phi_1, \phi_2, \ldots,
\phi_n)$ for the given $n$ and any $k\leq n$ is algebraically
independent. It follows that the coefficients
$\alpha^{(n)}_{k_1,k_2,\ldots,k_l}$ such that
$k_1+k_2+\ldots+k_l\leq n$ are uniquely defined. We assume that
the Taylor expansion of the function ${\cal P}^{(n)}(\phi_1,
\phi_2, \ldots, \phi_n)$ is divergent, therefore for $n$ large
enough the coefficients $\alpha^{(n)}_{k_1,k_2,\ldots,k_l}$ with
$k_1+k_2+\ldots+k_l > n$ are negligible, which justifies our
treatment of all the polynomials $E^{(k)}(\phi_1, \phi_2, \ldots,
\phi_n)$ as algebraically independent. In the limit of
$n\to\infty$ our treatment is strict.

Let us start with finding the solution such that
$\alpha^{(n)}_{k_1,k_2,\ldots,k_l}=0$ for $l\geq 2$. In this case
the invariant \eqref{expansion} reduces to:
\begin{equation}
\label{expansion2} {\cal P}^{(n)}(\phi_1, \phi_2, \ldots, \phi_n)
= \sum_{k=0}^\infty \alpha^{(n)}_k
E^{(k)}(\phi_1,\phi_2,\ldots,\phi_n).
\end{equation}
Inputting this expression into the condition
\eqref{axiom-probability} yields:
\begin{equation}
\begin{split}
\label{singleE} & \sum_{k=0}^\infty \alpha^{(n)}_k
E^{(k)}(\phi_1,\phi_2,\ldots,\phi_n) \sum_{s=0}^\infty
\alpha^{(m)}_s E^{(s)}(\xi_1,\xi_2,\ldots,\xi_m) \\
& = \sum_{t=0}^\infty \alpha^{(nm)}_t
E^{(t)}(\phi_1+\xi_1,\phi_1+\xi_2,\ldots,\phi_n+\xi_m).
\end{split}
\end{equation}
Using the definition of $E^{(n)}$ and Newton's formula we obtain:
\begin{equation}
\begin{split}
& E^{(t)}(\phi_1+\xi_1,\phi_1+\xi_2,\ldots,\phi_n+\xi_m) \\
& = \sum_{r=0}^t\binom{t}{r}E^{(r)}(\phi_1,\phi_2,\ldots,\phi_n)
E^{(t-r)}(\xi_1,\xi_2,\ldots,\xi_m).
\end{split}
\end{equation}
Inserting the above relation into \eqref{singleE} and using mutual
independence of the polynomials $E^{(k)}$ we obtain the condition
for the coefficients $\alpha^{(n)}_k$:
\begin{equation}
\label{Cauchy3} k!s!\alpha^{(n)}_k \alpha^{(m)}_s =
(k+s)!\alpha^{(nm)}_{k+s},
\end{equation}
which is the Cauchy equation with the following solution:
\begin{equation}
\alpha^{(n)}_k = \frac{1}{n^{\mbox{\scriptsize \em
\c{A}}}}\frac{\alpha^k}{k!},
\end{equation}
where $\alpha$ and {\em \c{A}} are arbitrary constants. Putting
this into the equation \eqref{expansion2} we obtain:
\begin{equation}
\label{expansion3}
\begin{split}
& {\cal P}^{(n)}(\phi_1, \phi_2,\ldots, \phi_n) =
\frac{1}{n^{\mbox{\scriptsize \em \c{A}}}}\sum_{k=0}^\infty
\frac{\alpha^k}{k!} E^{(k)}(\phi_1,\phi_2,\ldots,\phi_n) \\
& = \frac{1}{n^{\mbox{\scriptsize \em \c{A}}}}\left(e^{\alpha
\phi_1} + e^{\alpha \phi_2} +\ldots + e^{\alpha \phi_n}\right).
\end{split}
\end{equation}
Let us try to find out if the above special case generates all
possible solutions, or there are other irreducible functions
obeying the axioms \eqref{axiom-symmetry} and
\eqref{axiom-probability}. Consider the case of
$\alpha^{(n)}_{k_1,k2,\ldots,k_l} = 0$ for $l > N$ in
\eqref{expansion}. In this case we have:
\begin{equation}
\label{expansion4}
\begin{split}
{\cal P}^{(n)}(\phi_1, \phi_2, \ldots, \phi_n) = \sum_{k_1, k_2,
\ldots,k_N=0}^\infty \alpha^{(n)}_{k_1, k_2, \ldots,k_N} \\
E^{(k_1)}(\phi_1,\phi_2,\ldots,\phi_n)\cdots
E^{(k_N)}(\phi_1,\phi_2,\ldots,\phi_n).
\end{split}
\end{equation}
By substituting this into the axiom \eqref{axiom-probability} we
obtain the following condition:
\begin{widetext}
\begin{equation}
\sum_{\sigma,\sigma'}\alpha^{(n)}_{k_{\sigma(1)},\ldots,k_{\sigma(N)}}
\alpha^{(m)}_{s_{\sigma'(1)},\ldots,s_{\sigma'(N)}} =
\sum_{\pi,\pi'} \binom{k_{\pi(1)}+s_{\pi'(1)}}{k_{\pi(1)}}\cdots
\binom{k_{\pi(N)}+s_{\pi'(N)}}{k_{\pi(N)}}\,
\alpha^{(nm)}_{k_{\pi(1)}+s_{\pi'(1)},\ldots,k_{\pi(N)}+s_{\pi'(N)}},
\end{equation}
where $\sigma$, $\sigma'$, $\pi$, and $\pi'$ are arbitrary
permutations of an $N$-element set. Without a loss of generality
we can assume that the coefficients $\alpha^{(n)}_{k_1, k_2,
\ldots,k_N}$ are completely symmetric functions of $k_i$, because
any nonsymmetric component does not contribute to the overall sum
\eqref{expansion4} anyway. This assumptions yields:
\begin{equation}
\label{Cauchy} N! k_1! k_2!\cdots k_N! s_1! s_2!\cdots s_N!\,
\alpha^{(n)}_{k_1,\ldots,k_N} \alpha^{(m)}_{s_1,\ldots,s_N} =
\sum_{\pi} (k_1+s_{\pi(1)})!\cdots (k_N+s_{\pi(N)})!\,
\alpha^{(nm)}_{k_1+s_{\pi(1)},\ldots,k_N+s_{\pi(N)}},
\end{equation}
\end{widetext}
with the following solution:
\begin{equation}
\label{alpha2} \alpha^{(n)}_{k_1,k_2,\ldots,k_N} =
\frac{1}{n^{\mbox{\scriptsize \em \c{A}}'}}
\frac{\sum_\pi\alpha_{1}^{k_{\pi(1)}}\alpha_{2}^{k_{\pi(2)}}\cdots
\alpha_{N}^{k_{\pi(N)}}}{N!k_1! k_2!\cdots k_N!},
\end{equation}
where $\alpha_1, \alpha_2, \ldots, \alpha_N$ are arbitrary
constants. Let us verify what is the unknown function obeying
axioms \eqref{axiom-symmetry} and \eqref{axiom-probability} by
putting \eqref{alpha2} into the equation \eqref{expansion4}:
\vspace{2cm}
\begin{widetext}
\begin{equation}
\begin{split}
{\cal P}^{(n)}(\phi_1, \phi_2, \ldots, \phi_n) & =
\frac{1}{n^{\mbox{\scriptsize \em \c{A}}'}} \sum_{k_1, k_2,
\ldots,k_N=0}^\infty
\frac{\sum_\pi\alpha_{1}^{k_{\pi(1)}}\alpha_{2}^{k_{\pi(2)}}\cdots
\alpha_{N}^{k_{\pi(N)}}}{N!k_1! k_2!\cdots k_N!}
E^{(k_1)}(\phi_1,\phi_2,\ldots,\phi_n)\cdots
E^{(k_N)}(\phi_1,\phi_2,\ldots,\phi_n) \\
& = \frac{1}{n^{\mbox{\scriptsize \em \c{A}}'}} \sum_{k_1, k_2,
\ldots,k_N=0}^\infty \frac{\alpha_1^{k_1}\alpha_2^{k_2}\cdots
\alpha_N^{k_N}}{k_1! k_2!\cdots k_N!}
E^{(k_1)}(\phi_1,\phi_2,\ldots,\phi_n)\cdots
E^{(k_N)}(\phi_1,\phi_2,\ldots,\phi_n) \\
& = \frac{1}{n^{\mbox{\scriptsize \em \c{A}}'}}
\left(e^{\alpha_1\phi_1}+e^{\alpha_1\phi_2}+\ldots+e^{\alpha_1\phi_n}\right)\cdots
\left(e^{\alpha_N\phi_1}+e^{\alpha_N\phi_2}+\ldots+e^{\alpha_N\phi_n}\right).
\end{split}
\end{equation}
\end{widetext}
This shows that the only special case obeying the given axioms and
generating the general solution of the problem is given by the
expression \eqref{expansion3}. To complete the proof we notice
that the axiom \eqref{axiom-inverse} demands to take into account
only the products of pairs of solutions \eqref{expansion3} with
opposite signs $\alpha$ and $-\alpha$, as shown in the formula
\eqref{AEfunction}.

\end{appendix}


\begin{thebibliography}{99}

\bibitem{Bell1964}
J. S. Bell, Physics {\bf 1}, 195 (1964).

\bibitem{Frank1911}
Considerations based on similar assumptions, but in a much more
complicated form appeared for the first time in W. Ignatowsky,
Arch. Math. Phys {\bf 17} (1910) and P. Frank and H. Rothe, Ann.
der Phys. {\bf 34}, 825 (1911); the presented reasoning is a
modified version of the derivation by A. Szymacha, in {\em
Przestrzeń i ruch} (Wydawnictwo UW, Warsaw 1997).

\bibitem{Machildon1983}
L. Machildon, A. F. Antippa, and A. E. Everett, Can. J. Phys. {\bf
61}, 256 (1983); L. Machildon, A. F. Antippa, and A. E. Everett,
Phys. Rev. D {\bf 27}, 1740 (1983).

\bibitem{Cox1946}
R. T. Cox, Am. J. Phys. {\bf 14}, 1 (1946).

\bibitem{Wigner1939}
E. Wigner, Ann. Math. {\bf 40}, 149 (1939).

\bibitem{Feynman1949}
R. P. Feynman, Phys. Rev. {\bf 76}, 167 (1949).

\bibitem{Sierpinski1946}
W. Sierpi\'{n}ski, {\em Zasady Algebry Wy\.{z}szej}, (Warszawa
1946).

\bibitem{Aczel1966}
J. Acz\'{e}l, {\em Functional Equations and Their Applications},
(Academic Press, New York 1966).
\end{thebibliography}
\end{document}